\documentclass[12pt,preprint]{aastex}

\def\lya{\ifmmode {{\rm Ly}\alpha}\else
	Lyman-$\alpha$\fi}

\def\cm{\hbox{cm}}

\def\f17{f_{17}}
\def\lsun{L_\odot}
\def\Msun{M_\odot}

\def\kpc{\hbox{kpc}}
\def\pc{\hbox{pc}}

\def\kms{\hbox{\,km s$^{-1}$}}
\def\year{\hbox{yr}}
\def\Myr{\hbox{Myr}}

\def\arcsec{\ifmmode {''}\else{$''$}\fi}
\def\sqarcsec{\ifmmode{\square''}\else{$\square''$}\fi}

\def\kJy{\hbox{kJy}}

\def\ergcm2s{\ifmmode {\rm\,erg\,cm^{-2}\,s^{-1}}\else
                ${\rm\,ergs\,cm^{-2}\,s^{-1}}$\fi}
\def\ergsec{\ifmmode {\rm\,erg\,s^{-1}}\else
                ${\rm\,ergs\,s^{-1}}$\fi}
\def\kmsMpc{\ifmmode {\rm\,km\,s^{-1}\,Mpc^{-1}}\else
                ${\rm\,km\,s^{-1}\,Mpc^{-1}}$\fi}

\def\grapes3683{UDF~5225}
\def\mag{\,\hbox{mag}}

\begin{document}
\setcounter{footnote}{-1}
\title{A Redshift $z\approx 5.4$  \lya\ Emitting Galaxy with Linear 
Morphology in the GRAPES/UDF Field\thanks{
Accepted for publication in {\it The Astrophysical
Journal}, with tentative publication date March 2005}}

\author{
James E. Rhoads\altaffilmark{1,2},   
Nino Panagia\altaffilmark{1,3},      
Rogier A. Windhorst\altaffilmark{4},  
Sangeeta Malhotra\altaffilmark{1}, 
Norbert Pirzkal\altaffilmark{1},   
Chun Xu\altaffilmark{1},           
Louis Gregory Strolger\altaffilmark{1},  
Louis E. Bergeron\altaffilmark{1},  
Emanuele Daddi\altaffilmark{5},    
Harry Ferguson\altaffilmark{1},     
Jonathan P. Gardner\altaffilmark{6}, 
Caryl Gronwall\altaffilmark{7},      
Zoltan Haiman\altaffilmark{8},       
Anton Koekemoer\altaffilmark{1},     
Martin K\"{u}mmel\altaffilmark{12},  
Leonidas A. Moustakas\altaffilmark{1},  
Anna Pasquali\altaffilmark{9},       
Adam Riess\altaffilmark{1},          
Sperello di Serego Alighieri\altaffilmark{10},   
Massimo Stiavelli\altaffilmark{1},   
Zlatan Tsvetanov\altaffilmark{11},    
Joel Vernet\altaffilmark{10},         
Jeremy Walsh\altaffilmark{12},        
Haojing Yan\altaffilmark{13}
}

\begin{abstract}
We have discovered an extended \lya\ plume associated with a compact
source at redshift $z\approx 5.4$ in slitless spectroscopic data from
the Grism ACS Program for Extragalactic Science (GRAPES) project.  The
spatial extent of the emission is about $6 \times 1.5~\kpc$ ($1\arcsec
\times 0.25\arcsec$).  Combining our grism data and the broadband images
from the Hubble Ultra Deep Field (UDF) images, we find a
\lya\ line flux of $\sim 2.2 \times 10^{-17} \ergcm2s$ and surface
brightness $\sim 7 \times 10^{-17} \ergcm2s/\sqarcsec$.
The UDF images show
diffuse continuum emission associated with the \lya\ plume (hereafter
\grapes3683), with three embedded knots.  The morphology
of \grapes3683\ is highly suggestive of a galaxy in assembly.
It is moreover possible that the prominent \lya\ emission from this
object is due to an active nucleus, and that we are seeing the simultaneous
growth through accretion of a galaxy and its central black hole.  
Followup observations at higher spectral resolution could test this
hypothesis.
\end{abstract}

\keywords{galaxies: high redshift --- galaxies: formation --- 
galaxies: interactions --- galaxies: starburst --- 
galaxies: individual (\grapes3683)}

\altaffiltext{1}{Space Telescope Science Institute, 3700 San
 Martin Drive, Baltimore, MD 21218.}
\altaffiltext{2}{Email: rhoads@stsci.edu}
\altaffiltext{3}{Affiliated with the Space Telescope Division of the
European Space Agency, ESTEC, Noordwijk, the Netherlands}
\altaffiltext{4}{Arizona State University, Dept. of Physics \& Astronomy, 
P.O. Box 871504 Tempe, AZ 85287-1504, USA.  }
\altaffiltext{5}{European Southern Observatory, Garching, Germany.}
\altaffiltext{6}{Laboratory for Astronomy and Solar Physics, Code 681,
Goddard Space Flight Center, Greenbelt, MD 20771}
\altaffiltext{7}{The Pennsylvania State University, 525 Davey Lab,
University Park, PA 16802.}
\altaffiltext{8}{Columbia University, New York, NY.}
\altaffiltext{9}{Institute of Astronomy,
ETH Hoenggerberg, CH-8093 Zurich, Switzerland.}
\altaffiltext{10}{Osservatorio Astrofisico di Arcetri, Italy.}
\altaffiltext{11}{NASA Headquarters, Washington, DC}
\altaffiltext{12}{ST-ECF, Garching, Germany.}
\altaffiltext{13}{Spitzer Science Center, Caltech.}

\section{Introduction}
The Grism ACS Program for Extragalactic Science (GRAPES) 
project\footnote{HST GO program~9793, principal investigator S. Malhotra.} 
is a slitless spectroscopic survey that exploits the potential of 
the G800L grism on the Hubble Space Telescope's Advanced Camera for Surveys
(ACS) to achieve the most sensitive unbiased spectroscopy yet.  GRAPES is
targeted in the Hubble Ultra Deep Field (UDF) region, to complement
the UDF direct images, which are in turn the deepest optical imaging
to date (Beckwith et al. 2004).  The GRAPES survey, and in particular
our data analysis methods, are described in more detail by
Pirzkal et al. (2004).

One of the primary scientific goals of GRAPES is to study the
luminosity function of Lyman break galaxies (LBGs) using spectroscopically
confirmed samples at unprecedented sensitivity, and thereby to
constrain the faint end luminosity function slope.  We have begun this
effort through a targeted look at photometrically selected Lyman break
candidates, using both $i$-dropout galaxies from the UDF (Malhotra et
al 2004) and $V$-dropout galaxies which we have identified (Rhoads et al.
2004) following the selection criteria outlined by Giavalisco et al.
(2004b) in both the the v1.0 GOODS survey data (Giavalisco et al.
2004a) and the UDF.
(We refer to ACS and NICMOS filters by names of roughly corresponding
ground-based  filters: F435W $\rightarrow$ B; F606W $\rightarrow$ V;
F775W $\rightarrow$ i; F850LP $\rightarrow$ z; F110W $\rightarrow$ J; 
and F160W $\rightarrow$ H.)  Most of the confirmed Lyman break
objects are spatially compact, with sizes ($< 0.5\arcsec$) and morphologies
typical for the LBG population (e.g., Ferguson et al. 2004).  
In this {\it Letter}, we describe the
most prominent exception to this pattern we have encountered to date, a
$V$-dropout object (designated \grapes3683) that is exceptional in its
size, morphology, and spectroscopic properties.

Throughout this paper we use  the current concordance
cosmology ($H_0 = 71 \kmsMpc$, $\Omega_M = 0.27$, $\Omega_{total} = 1$;
see Spergel et al. 2003).  Magnitudes are given on the AB magnitude system,
so that magnitude zero corresponds to a flux density $f_\nu = 3.6 \kJy
= 3.6 \times 10^{-20} \ergcm2s {\hbox{Hz}^{-1}}$.
We present the observations in section~\ref{obssec}, compare them to
various physical models for \grapes3683\ in section~\ref{models},
and summarize our conclusions in section~\ref{theend}.

\section{Observational Properties of \grapes3683} \label{obssec}
\begin{figure}
\epsscale{0.75}
\plotone{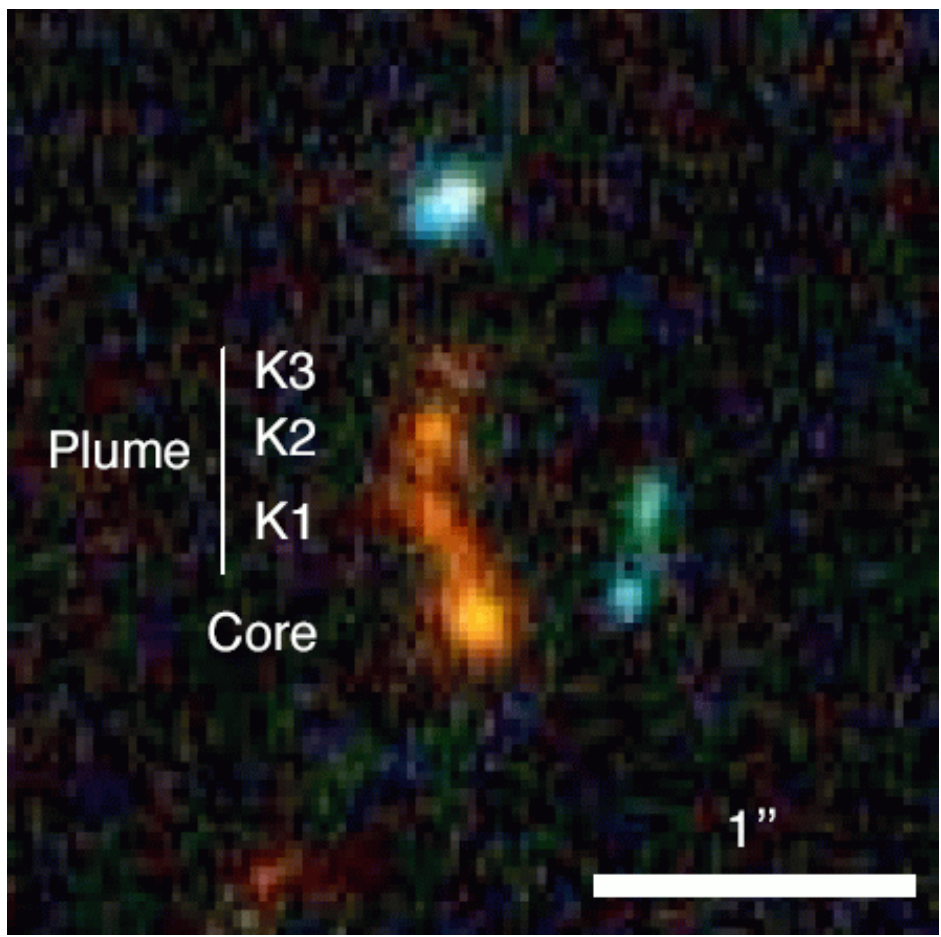}
\caption{Color composite image of \grapes3683\ from the Hubble
Ultra Deep Field data.  The elongated red source in the center is
\grapes3683.  We have labeled the ``core'' component at the bottom
and the ``plume'' extending towards the top of the image.  We also
label the three knots in the plume as ``K1,'' ``K2,'' and ``K3.''.
The length of the 
plume is nearly $1\arcsec$.  Blue colored objects nearby
are unrelated foreground sources.  Red represents the z filter, Green
an average of V and i, and blue the B filter.
Based on the color composite UDF image by Z. G. Levay.
\label{figcol}}
\end{figure}

\grapes3683\ has two morphological components (see fig.~\ref{figcol}):
A compact ``core,''
which is barely resolved in the ACS data (with full width
at half maximum $\approx 0.12\arcsec$ in the UDF z filter image,
which has a point spread function [PSF] with FWHM $\approx  
0.10\arcsec$)
and an extended ``plume'' with a size $\sim 1.0\arcsec \times 0.3\arcsec$.
There are three distinct condensations or ``knots'' within the plume.
The optical and near-IR colors of the object (discussed below) 
identify it as a Lyman break object, with an intrinsically blue
spectrum suppressed by the intergalactic medium at V band and bluer
wavelengths.

The size is unusual relative to $z\approx 5$ Lyman break galaxies selected 
photometrically from the GOODS data (Ferguson et al. 2004), which
show a broad peak between 0.1\arcsec\ and 0.5\arcsec\ (comparable to
the minor axis size of \grapes3683) and no galaxies as large
as 1\arcsec\ (the major axis size of \grapes3683).  \grapes3683\ itself
is not in the Ferguson et al. sample, being slightly fainter than
their flux limit.

\begin{figure}[ht]
\epsscale{0.75}
\plotone{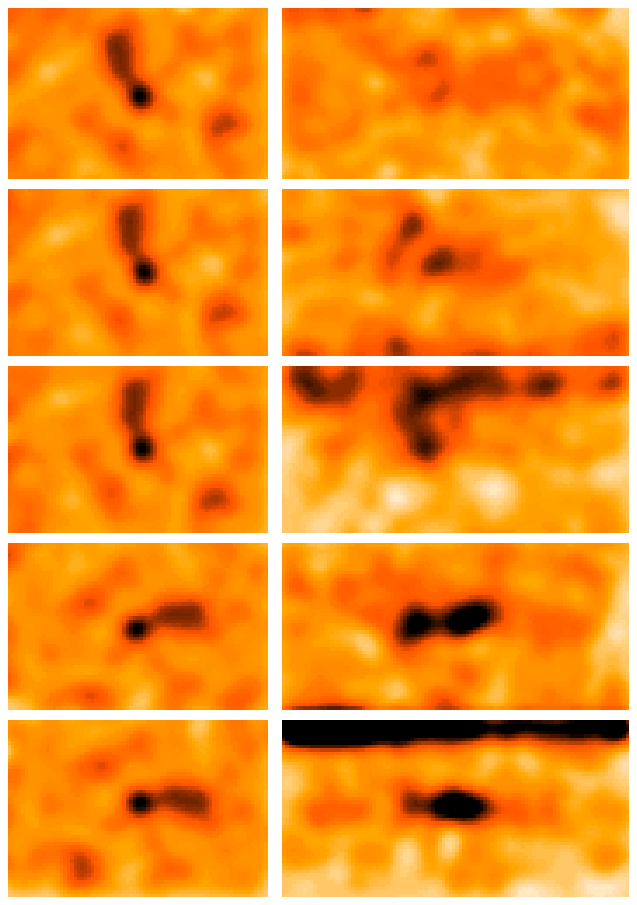}
\caption{The GOODS survey v1.0 direct image of \grapes3683\ in the z' (F850LP)
filter (left panels), together with the 2-D GRAPES spectra at each of five 
roll angles (right panels).
From top to bottom,  the angles shown are PA~117, 126, 134, 217, 
and~231$^\circ$. The direct image is the same
in all rows but is rotated to match the roll angle of each 2D grism
spectrum cutout.
All images have been smoothed with a Gaussian filter matched to the
angular size of the plume component ($0.3\arcsec$ FWHM) for maximum clarity.
The upper edge of the grism spectrum in the third PA shows contamination 
by an unrelated source elsewhere in the image.
The \lya\ emission from both plume and core is most clearly 
visible in the second and third panels; the first panel is at a similar
roll angle but is shallower.  In the fourth and fifth
panels, the line emission from the plume falls atop continuum
emission from the core, resulting in a higher surface brightness
in the dispersed image.
\label{fig2d}}
\end{figure}

The 2D ACS grism spectra of \grapes3683\ are shown in fig.~\ref{fig2d}.
They detect \grapes3683\ significantly in each of the
five epochs analyzed (see Pirzkal et al. 2004, and see also
Riess et al. 2004 for more detail on epoch~0).
Our strategy of using many roll angles results in a clean
separation of the core and plume spectra for 
epochs 0--2 ($PA=117^\circ$, 126$^\circ$, and 134$^\circ$, where
``$PA$'' refers to the position angle of the Hubble Space Telescope's
V3 axis),
while the spectra from the two components are superposed
in epochs 3 and 4 ($PA=217^\circ$ and  $231^\circ$).  
The spectrum from the northwestern tip of the plume is contaminated by
the spectrum of an unrelated, brighter object in the $PA=134^\circ$ data,
but otherwise the \grapes3683\ spectra are free of significant overlap.

Where the plume's spectrum can be examined independently of
other sources, including the core (i.e. epochs 0, 1, and in part 2),
it is dominated by a single
strong emission line at $\approx 7800$\AA.  Because this line falls
in the $i$ filter bandpass, and the plume is detected in both $i$
and $z$ images, we know that there must also be weak continuum emission
from the plume on the red side of the line.
The core shows both a break and a line at the same wavelength.
When both core and plume component spectra
are superposed (epochs 3--4), their combined line and continuum
flux results in a stronger spectroscopic detection.
We identify the line and break with \lya, based on their wavelength
coincidence in the spectrum of the core and based on the $B$-band
nondetection and very weak $V$-band flux of the source.  This then implies
a redshift $z\approx 5.42$, with an estimated uncertainty $\delta z \approx
0.07$. The object is near the upper end 
of the redshift range for V dropouts (and approaches the redshift range
of i dropouts).  

We measured the broadband optical magnitudes of the core and plume
components using the UDF images.  We defined apertures to match the
morphology of the core and plume components. These apertures follow
isophotes in a version of the UDF i band image smoothed with a
$0.14\arcsec$ FWHM Gaussian kernel, except at the boundary between the two
components.  There is no deep minimum in surface brightness separating
the core from the plume.  We find for the core $i = 27.94 \pm 0.017$,
$i-z= 0.51 \pm 0.025$, and $V-i = 2.3 \pm 0.12 \mag$, while for the
plume we obtain $i=27.39 \pm 0.023\mag$, $i-z=0.45 \pm 0.034\mag$, and
$V-i = 3.3 \pm 0.45\mag$.  Comparing our $i$-band flux with the flux
in the UDF catalog from STScI, we find that the masks contain about
55\% of the overall flux of the source, with the remaining flux in
lower surface brightness regions that we did not use for our color
measurements.  We did not attempt to measure the colors of the knots
within the plume individually, but inspection of figure~1 shows that
they are consistent with the overall colors of the plume.  Comparing
the colors of the main components, we see that the core is slightly
redder in $i-z$ and considerably bluer in $V-i$, but that the
significance of both statements is low: About $1.5\sigma$ for $i-z$
and $3\sigma$ for $V-i$.  (Note, we calculated the significance of the
$V-i$ color difference directly from flux ratios rather than
magnitudes.)  The $V-i$ colors are consistent with transmission
through the \lya\ forest of photons emitted at $912{\rm \AA} <
\lambda_{\rm rest} < 1215{\rm \AA}$.  For a flat ($f_\nu\approx
\hbox{constant}$) continuum, we would expect the \lya\ forest to
attenuate the V band flux by a factor of 14 for a source at redshift
$z=5.4$, based on the formalism of Madau (1995).

To eliminate any possibility that \grapes3683\ is simply a faint,
intrinsically red galaxy at a lower redshift, we also examined the
near-infrared colors of the object, using NICMOS images of the UDF (PI
R. Thompson).  We find a blue color ($z-J = 0.04$ and $J-H = -0.27$).
The NICMOS measurements were performed using an elliptical aperture
large enough to encompass the entire object (size $\approx 1.0\arcsec
\times 0.25\arcsec$).  We did not attempt separate IR measurements of
core and plume colors, given the larger pixel size, point spread
function size, and lower sensitivity of the UDF NICMOS data.  The
observed NIR colors support the identification of \grapes3683\ as a
Lyman break object with an intrinsically blue spectrum truncated below
\lya\ by the intergalactic medium.

Because \grapes3683\ is both faint and extended, it is near the
practical detection limit of the GRAPES data.  We therefore combine
the UDF imaging data, the NICMOS UDF observations, and the GRAPES
redshift to estimate the line flux and equivalent width by fitting the
spectral energy distribution (SED).  We model the intrinsic spectrum
as a power law continuum plus an unresolved \lya\ emission line,
modifying both by the \lya\ forest transmission calculated under the
model by Madau (1995).  The $z-J$ and $J-H$ colors are unaffected 
by \lya\ emission and  \lya\ forest absorption, so we use them
to constrain the intrinsic spectral slope to $\alpha \approx -2.4 
\pm 0.3$, where $f_\lambda \propto \lambda^{\alpha}$.
With the slope fixed, the $i-z$ and $V-i$ colors are determined by 
the line flux and the redshift (which determines how strongly 
\lya\ forest absorption affects broad band fluxes).  
For $z=5.42 \pm 0.07$, we find a rest frame equivalent width of
$70 \pm 30$\AA\ and an observed line flux of $(2.2 \pm 0.8) 
\times 10^{-17} \ergcm2s$.  The range in line flux and equivalent
width is primarily determined by the range of acceptable redshifts.
Changing the continuum slope within its plausible range has a
rather smaller effect on the line flux needed to match the
$i-z$ color.  The observed line flux corresponds to an approximate
surface brightness of $7\times 10^{-17}  \ergcm2s/\sqarcsec$.

To test whether the \lya\ emission in \grapes3683\ could be powered by
an active galactic nucleus (AGN), we examined the UDF direct images from
multiple epochs for variability.  We stacked the UDF $i$- and $z$-band
data into eight epochs of approximately equal exposure time.  We 
also stacked archival $z$-band imaging from HST program 9352, which
is shallower than 1/8 of the UDF data but gives a longer time
baseline ($\sim 1 \year$).
Subtracting the mean of the UDF stacks from each individual epoch
shows no significant residuals at the location of \grapes3683, leading
us to conclude that the source is not significantly variable in the
UDF data, with an upper limit to flux variations of $\la 10\%$. 
Unfortunately, at $z\approx 5.4$, any C{\scriptsize IV} line
lies beyond the wavelength coverage of the grism (as do all redder AGN
lines), while the N{\scriptsize V} line is not separable from the
\lya\ line at the resolution of the grism.
\grapes3683\ is not detected 
in X-rays, based on the 1.0 Msec Chandra Deep Field South catalog 
(Giacconi et al. 2002; Alexander et al 2003), 
though given its high redshift, this does not strongly
exclude an AGN: Only one AGN at $z>5$ has been discovered in the two
Chandra deep fields so far (Barger et al. 2002),
and an X-ray detection would only be expected for a rest frame hard-band
luminosity $> 1.8 \times 10^{43} \ergsec$.
Higher resolution optical spectra of \grapes3683\ could convincingly
determine whether or not it harbors an AGN by measuring the 
velocity width of the \lya\ emission.

\section{Models for \grapes3683} \label{models}
What is the nature of the \lya\ plume in \grapes3683?
The line luminosity and the equivalent width are both broadly
similar to those seen in narrowband-selected \lya\
samples (e.g., Rhoads et al. 2000, Malhotra \& Rhoads 2002,
Ouchi et al. 2003, Hu et al. 2004).  The morphology
is reminiscent of radio galaxies (e.g. Windhorst et al. 1998),
quasar jets, the ``\lya\ blobs'' that have been observed
at $z\sim 3$ (e.g., Steidel et al. 2000), and some 
other high redshift galaxies (Pascarelle et
al 1996,  Bunker et al. 2000, Keel et al. 2002).

We can consider several possibilities: A
recombination nebula powered by {\it in situ} star formation, possibly
triggered by a galactic merger; light from the 
core component scattered by either electrons or dust; or a 
a recombination nebula powered by the core.

\paragraph{{\it In situ} star formation:} \label{insitu}
Star formation would provide a local source of ionizing photons
within the plume.  It also produces lower-energy photons
that would dominate the rest-UV continuum redward of \lya.
The resulting continuum slope is the bluest among the models considered
here.  The rest frame \lya\ equivalent width for star formation should be
$\la 240$\AA\ based on stellar population models at Solar
metallicity (Charlot \& Fall 1993).
While observations of \lya\ galaxies at $z\approx 4.5$ often show larger
values (Malhotra \& Rhoads 2002), the presence of stars {\it in situ} will
always reduce the equivalent width relative to recombination models where
the ionizing photon source is distant.
Star formation is consistent with essentially any morphology.
A conversion factor of $1 \Msun/\year = 10^{42} \ergsec$ is widely used
for high redshift \lya\ emission, and would imply star formation
at $\ga 7 \Msun/\year$ in the plume, subject to the standard (but
untested) assumptions that the star formation follows a Kennicutt
(1983) initial mass function (IMF) and that case~B recombination is valid.
If we base our estimate instead on the  rest frame UV continuum emission
at $1425$\AA, as measured by the z band image, we find a star
formation rate of $\approx 4 \Msun/\year$ for the plume (plus
another $2.5 \Msun/\year$ in the core, assuming
the core light is not dominated by an active nucleus).
The conversion factors for both continuum and line
light are substantially uncertain, due to assumptions about the IMF
and about the effects of gas and dust on \lya\ radiative transfer.  
Thus, this constitutes remarkably good agreement.

\paragraph{Scattered Light:} If UV radiation from the core
encounters a sufficiently dense scattering medium, a detectable scattering
cone could be produced.  This is the least plausible explanation for
the \grapes3683\ plume, because the core luminosity would have to
be very large to power electron scattering, while dust scattering
would be unlikely to produce a large \lya\ equivalent width and a
blue continuum color.

We estimate the mass of scattering material required under the
approximation that the plume is a cone with length $6\,\kpc$ and base
diameter $\sim 2 \kpc$.  The corresponding volume (in physical,
not comoving, units) is $2\times 10^{65}
\cm^3$.  We estimate the mass for a scattering optical depth $\tau_e$
from the core to the end of the cone.  Then for electron scattering
(with $\sigma_T = 1.6 \times 10^{-24} \cm^{2}$) we find a number density
of $n_e \approx 25 \tau_e$ and a corresponding mass of
$\sim 5 \times 10^{9} \tau_e \Msun$ in the cone.  This 
gas would emit copious \lya\ radiation.  Indeed, to avoid producing more
than the observed \lya\ luminosity would require that $\tau_e \la 0.025$. 
Clumping of the scattering gas would be required to reproduce the
observed knots and the $30^\circ$ bend in the plume.  This clumping
would further enhance recombinations
and further reduce the maximum $\tau_e$ for electron scattering to dominate
the plume emission.

Such low optical depths imply that electron scattering is a very
inefficient way of producing the \lya\ nebulosity, because scattered
light will be suppressed by a factor $\tau_e$.  The scattering region
must then be illuminated at a much higher intensity than one would
na\"{\i}vely infer from the observed core flux.  Consider a simple toy
model where the central source has two emission components, one with
a $\sim 30^\circ$ opening angle that powers the observed plume, and one
isotropic that powers the observed core and that we allow to be
attenuated by optical depth $\tau_{abs}$ of absorption. To reproduce
the observed plume to core flux ratio, the collimated component would
then need to contain a fraction $\sim 1 / \left[ 1 + 0.6 \tau_e
\exp({\tau_{abs}}) \right]$, which becomes $\sim 1 - 0.6 \tau_e \ga
98\%$ in the limit where $\tau_{abs} \la 1$.  This would correspond to
a total source UV luminosity $\nu L_\nu \ga 3\times 10^{45} \ergsec$
(measured in the $z$-band, which is $\lambda_{\rm rest} \approx 1410$\AA)
i.e., $7\times
10^{11} \lsun$ or an angle-averaged absolute AB magnitude of $-23.7$.
This luminosity would increase if either the assumed $\tau_e$ or
degree of collimation were reduced.  Such a model would likely require 
an AGN in the core, because star light cannot be tightly collimated,
which would increase the luminosity requirement by another factor
of $\ga 10$ while requiring also $\tau_{abs} \ga 2$ on the line
of sight to the core.
The mass of ionized hydrogen required in an electron scattering
scenario would be modest, $\la 10^8 \Msun$.

Dust scattering in an ionized medium could work, as
long as the optical depth is suitably small so that (a) there is no
significant reddening of the scattered continuum light,
and (b) Ly-a radiation is not selectively absorbed relatively to the
continuum.  Condition (a) requires $\tau \la 0.1$ at $\lambda_{\rm rest}
\approx 1300$\AA, and
similarly condition (b) requires $\tau \la 0.05$ at  $\lambda_{\rm rest}
\approx 1216$\AA\ to avoid attenuation of \lya\ by more than a
factor of 2 (Panagia and Ranieri 1973a, b).  The associated total mass of
gas and dust would be 
$M_{\rm plume} \approx 10^7 \tau_{\rm dust}(1216{\rm \AA}) \Msun$ 
assuming a dust-to-gas mass ratio of $0.01$.  Inserting
$\tau_{\rm dust} <0.1$  gives  $M_{\hbox{plume}} \sim 10^6 \Msun$.
The requirement on the core luminosity can be derived exactly as for
the electron scattering case, but now with larger scattering opacity,
so the final constraint becomes $\nu L_\nu \ga 3 \times 10^{44} \ergsec$.

An additional concern with scattering models is that
the intensity of scattered light should
decrease away from the AGN (falling off as $r^{-2}$ in the simplest
case, though projection effects and anisotropic quasar emission could
modify this).  Yet the surface brightness of the plume is constant
to within factors of two over a factor of $\sim 8$ in distance from the
core, which would require the column density to increase with
distance as $\sim r^2$.

\paragraph{Recombination powered by core light:}
Objects like ``\lya\ blobs'' (e.g., Steidel et al. 2000) and
some radio galaxies have extensive \lya\ nebulae (up to $100 \kpc$ in 
size) that are likely powered by AGN emission.  We now consider whether
\grapes3683\ could be a physically similar object on a smaller scale.

Such a model is similar to the electron scattering model but with higher
densities, where recombinations become more important than electron
scattering.  In this case the mass of ionized hydrogen involved
would be $\ga 10^8 \Msun$.  The core luminosity could be much
lower in this case: For an ionization bounded nebula, it need only
be about twice the measured \lya\ luminosity, i.e., $> 2\times 10^{43}
\ergsec$ in ionizing radiation, although it could be much larger
if the radiation is largely isotropic and the morphology of the
plume is set by where there is substantial gas.

However, in this case, the continuum emission from the plume would be pure 
nebular emission, with components from the two-photon process,
bound-free emission, and free-free emission.  The expected equivalent
width of the line would then be $> 1000$\AA\ (rest frame) and the
z band continuum should be much weaker than we see.  While dust
attenuation of the \lya\ might help, the model would have to be fairly
contrived to simultaneously fit the \lya\ flux, continuum flux, 
equivalent width, and color in the plume, and we therefore disfavor
this model also.  The only countervailing argument comes
from the V band flux, which should be absent in a two-photon continuum
and is indeed rather weaker in the plume than the core.  However none
of the V band detections are strong, so it is not clear if the
difference in $V-i$ and $V-z$ colors between the two components is 
significant.

\paragraph{A Merger Scenario:}
Consider now the possibility that  \grapes3683\ is a major
merger in progress at $z\approx5.4$.   The
nuclei of the two interacting galaxies would then presumably
be the ``core'' component we have discussed plus the brightest
``knot'' in the plume, which lies $0.63\arcsec$ away.
Quantitative morphological tests based on the asymmetry parameter
$A$ (e.g., Conselice et al. 2003) are consistent with this
scenario, based on an analysis of GRAPES object morphologies
now in progress (Pirzkal et al. 2004b).
If we measure asymmetry relative to the center of
light for \grapes3683\ (i.e., almost half way from the core to
the end of the plume), we find $A(z) = 0.34 \pm 0.09$ and
$A(i) = 0.16 \pm 0.09$ for the $z$ and $i$ filters.  If we instead
measure $A$ with the center placed in the core component, we get
much higher numbers, $A(z) = 0.45 \pm 0.13$ and $A(i) = 0.43 \pm 0.13$.
Mergers are typically found to have $A\ga 0.35$ (Conselice et al. 2003).
Further discussion of these parameters for the full $i$- and $V$-drop 
GRAPES samples will be presented in Pirzkal et al (2004b).

The projected separation of the core and knot, $\sim 4\kpc$, would then imply a
crossing time of order $40 v_{100} \Myr$, where $v_{100} \sim 1$ is
the relative velocity of the two components in units of $100 \kms$.
Multiplying this by the star formation rates inferred in 
section~\ref{insitu} implies formation of some few $\times 10^8 \Msun$
of stars in the course of the interaction.  

The star formation rate per unit area, based conservatively on
the UV-derived SFR for the plume, is $\sim 0.4 \Msun \year^{-1} \kpc^{-2}$.
Comparing to the global Schmidt law for star formation, we would
infer a gas mass surface density of $\sim 800 \Msun \pc^{-2}$, for
a total gas mass of order $7 \times 10^9\Msun$.  This number implies
a gas consumption time scale of $\sim 10 t_{dyn} \sim 600 \Myr$, though
if  star formation is proceeding at an atypically high rate (driven by
interaction), the gas reservoir could be smaller.

\section{Discussion} \label{theend}
We have examined several possible scenarios for the observed
properties of the galaxy \grapes3683, a very faint and high
redshift object with a core-plume morphology and prominent
\lya\ emission.  We conclude that the \lya\ emission is
most likely powered by {\it in situ\/} star formation throughout
the object.  Present evidence neither requires nor rules out 
the presence of an  AGN in the core component.  Followup spectroscopy
using higher spectral resolution and/or coverage into the near infrared
would provide new information that could settle the AGN question.
Tidally triggered star formation in a merging galaxy pair seems to describe
the galaxy well.  As such, it may be a particularly spectacular example
of the broad class of star-forming Lyman break galaxies that 
dominate the galaxy population observed in the Hubble Ultra Deep Field
and other very sensitive, high redshift galaxy surveys.

\acknowledgements
We thank the STScI Director and the UDF team for their hard work in
designing and executing the UDF experiment.
We thank Zoltan G. Levay for producing the color composite
shown in figure~\ref{figcol}.  
This work has been supported under grant number HST-GO-09793.
C.G. acknowledges support from NSF-AST-0137927.
L.A.M. acknowledges support by NASA through contract number 1224666
issued by the Jet Propulsion Laboratory, California Institute of
Technology under NASA contract 1407.

\end{document}